\documentclass[final,authoryear,5p,times,twocolumn]{elsarticle}
\usepackage{epsfig}
\usepackage{amssymb}
\journal{Journal of Theoretical Biology}

\begin{document}
\begin{frontmatter}

\title{Effectiveness of conditional punishment for the evolution of public cooperation}

\author[mfa]{Attila Szolnoki}
\author[umb]{Matja{\v z} Perc}

\address[mfa]{Institute of Technical Physics and Materials Science, Research Centre for Natural Sciences, Hungarian Academy of Sciences, P.O. Box 49, H-1525 Budapest, Hungary}
\address[umb]{Faculty of Natural Sciences and Mathematics, University of Maribor, Koro{\v s}ka cesta 160, SI-2000 Maribor, Slovenia}

\begin{abstract}
Collective actions, from city marathons to labor strikes, are often mass-driven and subject to the snowball effect. Motivated by this, we study evolutionary advantages of conditional punishment in the spatial public goods game. Unlike unconditional punishers who always impose the same fines on defectors, conditional punishers do so proportionally with the number of other punishers in the group. Phase diagrams in dependence on the punishment fine and cost reveal that the two types of punishers cannot coexist. Spontaneous coarsening of the two strategies leads to an indirect territorial competition with the defectors, which is won by unconditional punishers only if the sanctioning is inexpensive. Otherwise conditional punishers are the victors of the indirect competition, indicating that under more realistic conditions they are indeed the more effective strategy. Both continuous and discontinuous phase transitions as well as tricritical points characterize the complex evolutionary dynamics, which is due to multipoint interactions that are introduced by conditional punishment. We propose indirect territorial competition as a generally applicable mechanism relying on pattern formation, by means of which spatial structure can be utilized by seemingly subordinate strategies to avoid evolutionary extinction.
\end{abstract}


\end{frontmatter}

\section{Introduction}
Adherence to law in human societies is maintained by sanctioning. If the law has been broken retribution will follow. The looming threat of punishment should thus avert us from engaging into illegal activities. Positive incentives for adhering to the law are much less common and restricted mainly to motivating past offenders to stay on the right path. The evolutionary origins of this are difficult to determine. Our other-regarding abilities are believed to have been ignited by challenges in rearing offspring that survived \citep{hrdy_11}, although there is also evidence suggesting that between-group conflicts may have been instrumental too \citep{bowles_11}. Both options are viable and suggest that members of human societies were initially more prone to adherence than they were to disobedience and disregard of expected behavioral patterns. Punishment may therefore have emerged spontaneously as a way of treating the minority that misbehaved. It would have been much more tedious and taxing to reward all those that behaved properly. And this might have eventually led to the evolution of the legal system that is in place today, rather than to the evolution of a large-scale rewarding system.

Although to cooperate is certainly not the same as adhering to the law, in the light of preceding arguments it may nevertheless be little surprising that punishment, much more so than rewarding, would be considered as the preferred method of choice for averting the threatening ``tragedy of the commons'' \citep{hardin_g_s68}. Past research has in fact focused more on punishment than reward for promoting and maintaining public cooperation \citep{clutton_brock_n95, fehr_n02, fehr_n03, semmann_n03, de-quervain_s04, fowler_pnas05, hauert_s07, gachter_s08, ohtsuki_n09, rockenbach_n09}, with the general conclusion being that sanctioning is indeed more effective \citep{sigmund_pnas01, sigmund_tee07}. Only recently, the efficiency of punishment has been brought into questioning. Rewarding appears to offer evolutionary advantages that go beyond those warranted by punishment \citep{dreber_n08, rand_s09, hilbe_prsb10, szolnoki_epl10, hauert_jtb10, szolnoki_njp12}, while the introduction of antisocial punishment might render the concept of sanctioning altogether ineffective \citep{herrmann_s08, rand_jtb10, rand_nc11, garcia_jtb12, hilbe_srep12}. Although it is beyond the scope of the present work to discuss the potential relevance and feasibility of these strategic choices, the continued appeal of punishment as a means to promote public cooperation, as evidenced by recent studies on institutionalized punishment \citep{sigmund_n10, szolnoki_pre11, perc_srep12, traulsen_prsb12}, the coevolution and self-organization of punishment \citep{boyd_s10, perc_njp12}, as well as its many variants \citep{mathew_pnas11, baldassarri_pnas11, sasaki_pnas12}, ought to attest to its lasting effectiveness and thus lend support to further explorations to that effect.

With this in mind, we here study potential evolutionary advantages of conditional punishment in the spatial public goods game \citep{wakano_pnas09, szolnoki_pre09c}. It is clear that spatial structure plays a pivotal role by the evolution of cooperation, as comprehensively reviewed in \citep{szabo_pr07, roca_plr09, perc_bs10}. Although recent large-scale human experiments indicate otherwise \citep{gracia-lazaro_srep12, gracia-lazaro_pnas12}, there is ample theoretical evidence indicating that relaxing the simplification of well-mixed interactions may lead to qualitatively different results that are due to pattern formation and intricate organization of the competing strategies, which reveals itself in most unexpected ways \citep{szolnoki_prl12}. The seminal paper introducing games on grids is due to Nowak and May \citep{nowak_n92b}, while recent works concerning the spatial public goods game have considered the relevance of complex interaction networks and coevolution \citep{lozano_ploso08, wu_t_epl09, wu_t_pre09, gomez-gardenes_c11, gomez-gardenes_epl11, pena_pone12}, diversity \citep{santos_n08, fort_epl08, wang_j_pre10b, santos_jtb12}, the risk of collective failures \citep{santos_pnas11, chen_xj_pre12b}, the appropriate partner and opponent selection \citep{wu_t_pre09, zhang_hf_epl11, brede_acs12}, the population density \citep{wang_z_pre12b}, conditional cooperation \citep{szolnoki_pre12}, heterogeneous wealth distributions \citep{wang_j_pre10b}, directed investments \citep{vukov_jtb11}, selection pressure \citep{van-segbroeck_njp11, pinheiro_njp12}, as well as both the joker \citep{arenas_jtb11,requejo_pre12b} and the Matthew effect \citep{perc_pre11}, to name but a few.

The relevance of structured populations for the success of punishment is also thoroughly documented \citep{brandt_prsb03, nakamaru_eer05, helbing_ploscb10, szolnoki_pre11, perc_njp12}. Since the number of competing strategies can be three \citep{hauert_s02, bowles_tpb04, brandt_pnas05, helbing_njp10}, four \citep{sigmund_pnas01, ohtsuki_n09}, or even higher \citep{henrich_jtb01, dreber_n08, rand_jtb10}, besides traditional cooperators and defectors taking into account also all the different forms of punishment, the simulations of spatial systems ought to be done with a lot of caution. If imitation governs the evolutionary process, which is certainly a reasonable assumption given that it has a positive impact even at weak selection pressure \citep{masuda_srep12, mobilia_pre12, szolnoki_srep12}, possible stable solutions of the whole system are all the solutions of each subsystem, comprising only a subset of all the original strategies \citep{szabo_pr07}. The most stable solution can only be determined by performing a systematic check of the direction of invasion between all possible pairs of subsystem solutions that are separated by an interface in the spatial system. Of course many of the subsystem solutions will not be stable, and along several of the interfaces the victor will be obvious, which may significantly reduce the complexity of the problem. Nevertheless, the belief that simulations of spatial games are subject to no restrictions in terms of the number of competing strategies is wrong and should not be perpetuated based on the few rare exceptions that considered prohibitively high numbers of competing strategies in spatial games but did not take properly into account the limitation and pitfalls, including accidental extinctions due to insufficiently large system size.

Here we consider four competing strategies on a square lattice. Cooperators, who contribute to the public good but abstain from punishing defectors are the second-order free-riders, and they can seriously challenge the success of sanctioning \citep{panchanathan_n04, fowler_n05b}. Defectors neither contribute to the public good nor to sanctioning. Notably, the impact of double moral behavior, i.e., defectors who punish other defectors, has been studied before in \citep{helbing_ploscb10}. Finally, we have conditional and unconditional punishers, who both contribute to the public good as well as to punishing defectors. However, while unconditional punishers always impose the maximal fine on defectors, conditional punishers fine defectors proportionally to the number of other punishers, either conditional or unconditional, in the group. Importantly, the cost of punishment that the punishers have to bear is always proportional with the imposed fine, so that the ratio between the fine and the cost is the same for both types of punishment. It can be argued that conditional punishers act according to the ``majority driven'' principle, which has in fact been confirmed experimentally for the severity of punishment in a public goods game setting \citep{kodaka_pone12}. Compared to the three-strategy game entailing only cooperators, defectors and unconditional punishers \citep{helbing_njp10}, we will show that the introduction of conditional punishers lowers the minimally required fine that is needed for cooperation to grab hold in the population, and that in the more relevant parameter space where the cost of sanctioning is comparable to the imposed fines, the conditional way of punishing is in fact more effective. Moreover, we will show that the indirect territorial competition reported first in \citep{helbing_ploscb10} can be observed also for other strategy pairs, and that in general it is responsible for discontinuous phase transitions between stable solutions of the game. We will extend and explain these results in detail in Section III, while now we proceed with a detailed description of the studied spatial public goods game with conditional punishment.

\section{Spatial public goods game with conditional punishment}
The public goods game is staged on a square lattice with periodic boundary conditions where $L^2$ players are arranged into overlapping groups of size $G=5$ such that everyone is connected to its $k=G-1$ nearest neighbors. Accordingly, each individual belongs to $g=1,\ldots,G$ different groups. Initially each player on site $x$ is designated either as a cooperator ($s_x = C$), defector ($s_x = D$), conditional punisher ($s_x = P_c$), or unconditional punisher ($s_x = P_u$) with equal probability. Except defectors, all three other strategies contribute a fixed amount, here considered being equal to $1$ without loss of generality, to the public good. The sum of all contributions in each group is multiplied by the synergy factor $r$ and the resulting public goods are distributed equally amongst all the group members irrespective of their contributions.

Punishment is taken into account as follows. Cooperators do not participate in the sanctioning of defectors, and hence become the second-order free-riders \citep{panchanathan_n04, fowler_n05b}.
An unconditional punisher imposes the fine $\beta/(G-1)$ on each defector within the group, and bears the related punishment cost $\gamma/(G-1)$, regardless of the presence of other strategies. According to this parametrization, a single defector is punished by the total fine $\beta$ in a homogeneous group of unconditional punishers. Conditional punishers, on the other hand, impose a fine and carry the cost that is proportional to the number of other punishers, either conditional or unconditional, within the group. For example, if a conditional punisher is surrounded solely by other defectors and pure cooperators, the fine imposed on each defector will be just $1/(G-1)$ of the maximal value. Importantly, the ratio between the imposed fine and the related cost is always the same, which is essential because otherwise the efficiency of unconditional and conditional punishment cannot be properly compared. Designating then the number of cooperators, defectors, conditional punishers and unconditional punishers within the group $g$ as $N_C$, $N_D$, $N_{P_c}$ and $N_{P_u}$, respectively, the payoffs of the four strategies stemming from this particular group $g$ are:
\begin{equation}
\pi_C^g=r \frac{N_C+N_{P_c}+N_{P_u}}{G} - 1, \nonumber
\label{piC}
\end{equation}
\begin{equation}
\pi_D^g=r \frac{N_C+N_{P_c}+N_{P_u}}{G} - N_{P_c} \frac{(N_{P_c}+N_{P_u}) \beta}{(G-1)^2} - N_{P_u} \frac{\beta}{G-1},
\label{piD}
\end{equation}
\begin{equation}
\pi_{P_u}^g=r \frac{N_C+N_{P_c}+N_{P_u}}{G} - 1 - N_D \frac{\gamma}{G-1},
\label{piPc}
\end{equation}
\begin{equation}
\pi_{P_c}^g=r \frac{N_C+N_{P_c}+N_{P_u}}{G} - 1 - N_D \frac{(N_{P_c}+N_{P_u}) \gamma}{(G-1)^2}.
\label{piPu}
\end{equation}
Notably, the first subtraction in Eq.~\ref{piD}, which determines the payoff of defectors, is due to conditional punishment, while the second one is due to unconditional punishment. The punishers  bear the additional costs accordingly, as described by the last terms in Eqs.~\ref{piPc} and \ref{piPu}. It is also worth emphasizing that conditional punishment introduces multi-point interactions in that the fine imposed on defectors as well as the related additional costs of conditional punishers cannot be derived simply from straightforward two-player interactions. Note that the fine imposed by player A onto player B depends not just on the strategies of these two players, but also on the strategies of other players within the group.

Monte Carlo simulations of the game are carried out comprising the following elementary steps. A randomly selected player $x$ plays the public goods game with its $k$ partners as a member of all the $g$ groups, whereby its overall payoff $\pi_{s_x}$ is thus the sum of all the payoffs acquired in the five groups. Next, player $x$ chooses one of its nearest neighbors at random, and the chosen co-player $y$ also acquires its payoff $\pi_{s_y}$ in the same way. Finally, player $x$ enforces its strategy $s_x$ onto player $y$ with a probability $q=1/\{1+\exp[(\pi_{s_y}-\pi_{s_x})/K]\}$, where $K=0.5$ quantifies the uncertainty by strategy adoptions \citep{szolnoki_pre09c}, implying that better performing players are readily adopted, although it is not impossible to adopt the strategy of a player performing worse. Such errors in decision making can be attributed to mistakes and external influences that adversely affect the evaluation of the opponent. Each Monte Carlo step (MCS) gives a chance for every player to enforce its strategy onto one of the neighbors once on average. The average densities of the four strategies were determined in the stationary state after sufficiently long relaxation times. Depending on the actual conditions, such as the proximity to phase transition points and the typical size of emerging spatial patterns, the linear system size was varied from $L=400$ to $3200$ and the relaxation time was varied from $10^5$ to $10^7$ MCS to ensure proper accuracy. In general, the application of larger system size was necessary to determine the accurate location of discontinuous phase transitions.

\section{Results}

\begin{figure}
\centerline{\epsfig{file=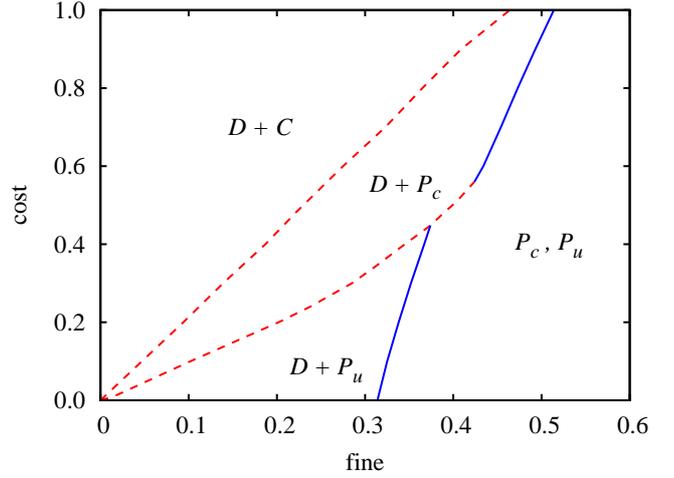,width=8.5cm}}
\caption{\label{phase38} Phase diagram, depicting the strategies ($C$ - cooperators, $D$ - defectors, $P_c$ - conditional punishers, $P_u$ - unconditional punishers) that remain on the square lattice in the stationary state at $r=3.8$, in dependence on the punishment fine $\beta$ and cost $\gamma$. Red dashed lines denote first-order discontinuous phase transitions, while solid blue lines denote second-order continuous phase transitions. At high fines $P_c$ and $P_u$ become neutral after cooperators and defectors die out (see main text for details).}
\end{figure}

For the classical two-strategy spatial public goods game that is contested solely between cooperators and defectors, there exists a critical value of $r$ above which cooperation is no longer possible. On the square lattice with overlapping groups containing five players each, the critical value is equal to $r=3.74$ at the applied value of $K$ \citep{szolnoki_pre09c}. Accordingly, it is of interest to investigate the impact of punishment above and below this threshold, as the presence of cooperators, which actually become the second-order free-riders because they abstain from punishing defectors \citep{panchanathan_n04, fowler_n05b}, is likely to affect the evolutionary outcome.

\begin{figure*}
\centerline{\epsfig{file=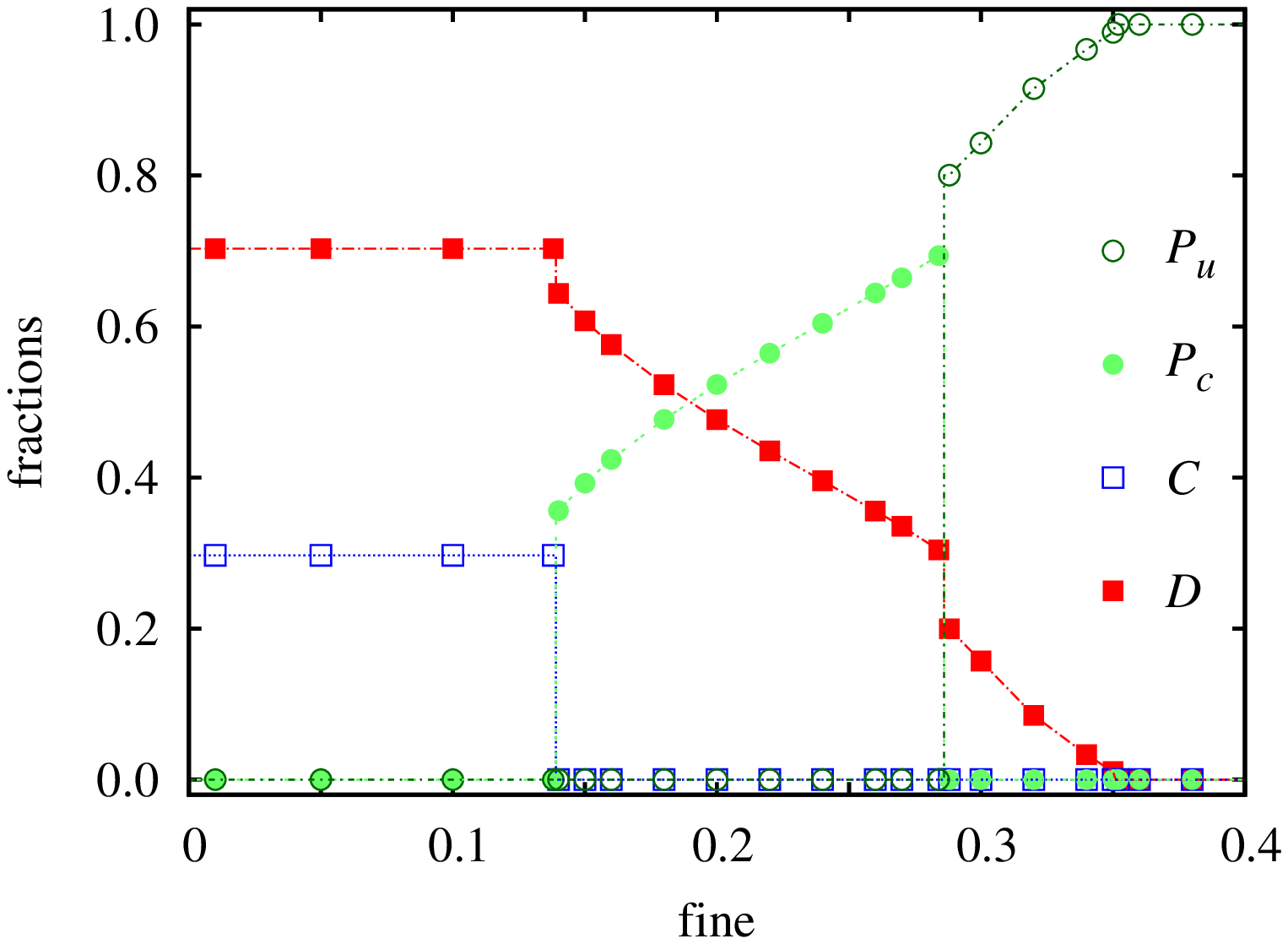,width=5.8cm}\epsfig{file=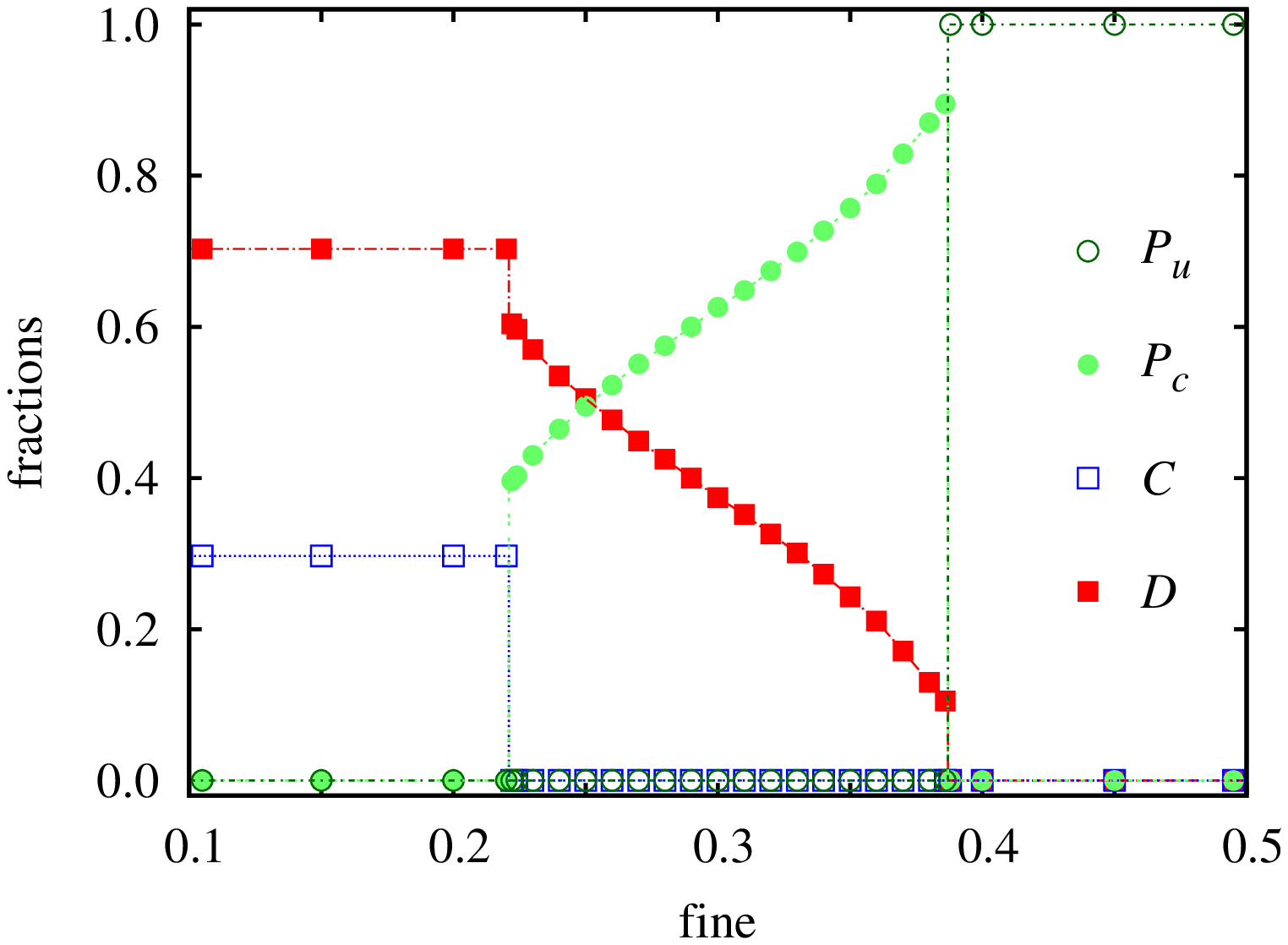,width=5.8cm}\epsfig{file=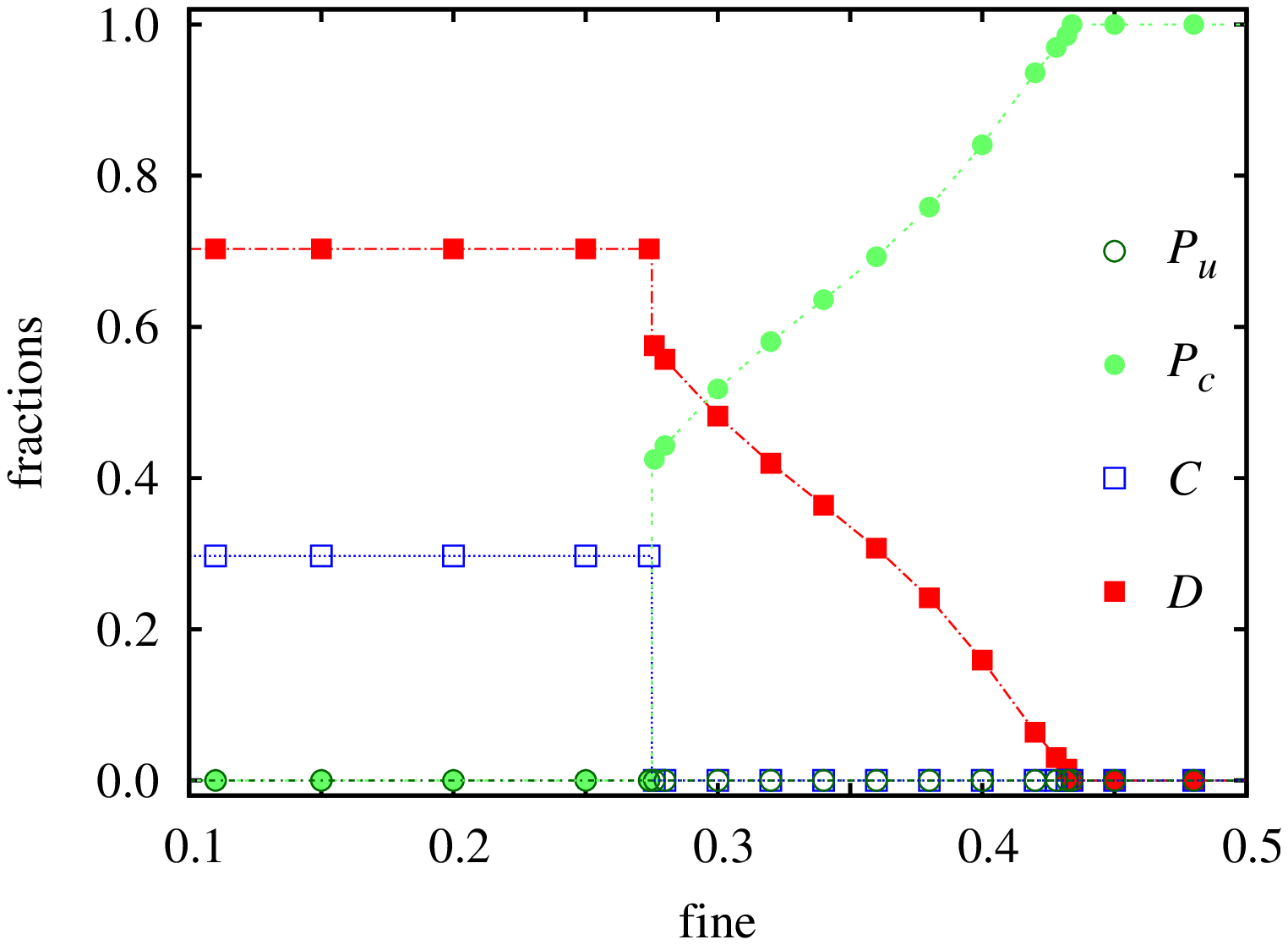,width=5.8cm}}
\caption{\label{cross38} Representative cross-sections of the phase diagram presented in Fig.~\ref{phase38}. Depicted are the stationary densities of the four competing strategies (see legend) in dependence on the punishment fine $\beta$, as obtained for three different values of cost: $\gamma=0.3$ (left), $\gamma=0.45$ (middle), and $\gamma=0.6$ (right).}
\end{figure*}

We begin by setting $r=3.8$, where cooperators alone are able to survive in the presence of defectors, and determine the survivability of the four competing strategies in dependence on the punishment fine $\beta$ and cost $\gamma$. The full $\beta-\gamma$ phase diagram is presented in Fig.~\ref{phase38}. It can be observed that the mixed $D+C$ phase dominates if only the ratio between the punishment cost and fine is sufficiently high. As soon as $\beta$ exceeds a threshold, the mixed $D+C$ phase gives way to a mixed $D+P_c$ phase via a first-order discontinuous phase transition. Naturally, the higher the cost of punishment, the larger the value of $\beta$ that is needed to evoke this transition. If we compare this phase diagram with the one obtained for the three-strategy public goods game that does not contain conditional punishers \citep{helbing_njp10}, we find that in the present case the phase transition line that delineates the punishment-free state is actually shifted to smaller fines. In other words, conditional punishers can subvert second-order free-riders even under less friendly conditions, i.e., when the punishment is more costly. If, however, the $\gamma/\beta$ ratio is sufficiently low, unconditional punishers are able to play out the advantage of higher fines, and accordingly they become the most successful in resisting the invasions of defectors. Again a first-order discontinuous phase transition delineates the $D+P_c$ phase and the $D+P_u$ phase. It is crucial to note that neither $P_c$ and $P_u$ nor $C$ and $P_u$ (or $P_c)$ are able to coexist. Spontaneous coarsening of these three strategies leads to an indirect territorial battle that is always mediated by defectors, and it is believed to be generally valid that this type of indirect competition between different strategies always leads to discontinuous phase transitions. Such an evolutionary dynamics has been first described in \citep{helbing_ploscb10}, while here we argue that it is indeed much more common than originally assumed, and that it may be a general mechanism relying on pattern formation, by means of which spatial structure can be exploited to create evolutionary advantages for seemingly inferior strategies (as is the case if comparing cooperators, i.e., second-order free-riders, and punishers in a well-mixed population). It is in fact by means of this mechanism that unconditional punishers are able to crowd out cooperators, and for still higher fines, the unconditional punishers are able to crowd out conditional punishers. Under special conditions, for $r=3.8$ given at $\beta \simeq 0.42$ and $\gamma \simeq 0.56$, the discontinuous and continuous transition lines join, which we conjecture to be a tricritical point. Above this point $P_u$ cannot survive, and hence the $D+P_c$ phase goes to the pure $P_c$ phase via a second-order continuous phase transition. When the imposed fines are even higher, both $D$ and $C$ die out, and from that point onwards $P_c$ and $P_u$ become neutral. According to the voter-type dynamics \citep{dornic_prl01}, a logarithmically slow coarsening determines the final state, which can be either a homogeneous $P_c$ or a homogeneous $P_u$ phase (hence the $P_c,P_u$ notation in Fig.~\ref{phase38}). The probability to reach either depends on the initial ratio of the two strategies at the time $D$ and $C$ die out. Accordingly, it is more likely that a homogenous $P_c$ phase will be reached at higher costs, while $P_u$ are likelier to dominate for values of $\gamma$ that are below the tricritical point.

\begin{figure}
\centerline{\epsfig{file=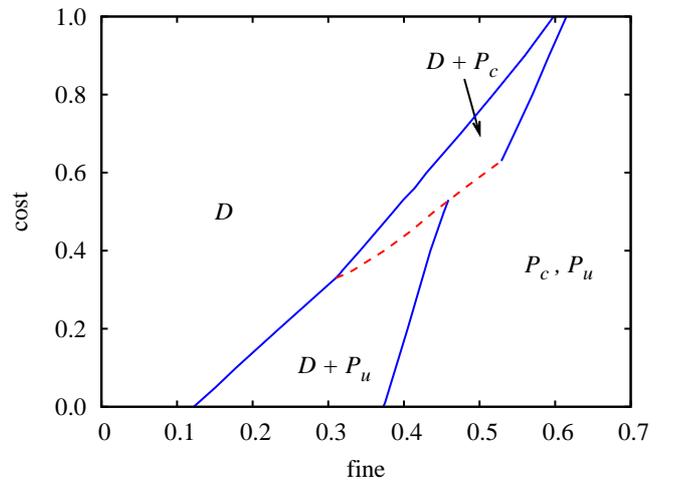,width=8.5cm}}
\caption{\label{phase35} Phase diagram, depicting the strategies that remain on the square lattice in the stationary state at $r=3.5$, in dependence on the punishment fine $\beta$ and cost $\gamma$. Note that at this value of $r$ cooperators are unable to survive alone in the presence of defectors \citep{szolnoki_pre09c}. Notation and line styles are the same as those used in Fig.~\ref{cross38}.}
\end{figure}

\begin{figure*}
\centerline{\epsfig{file=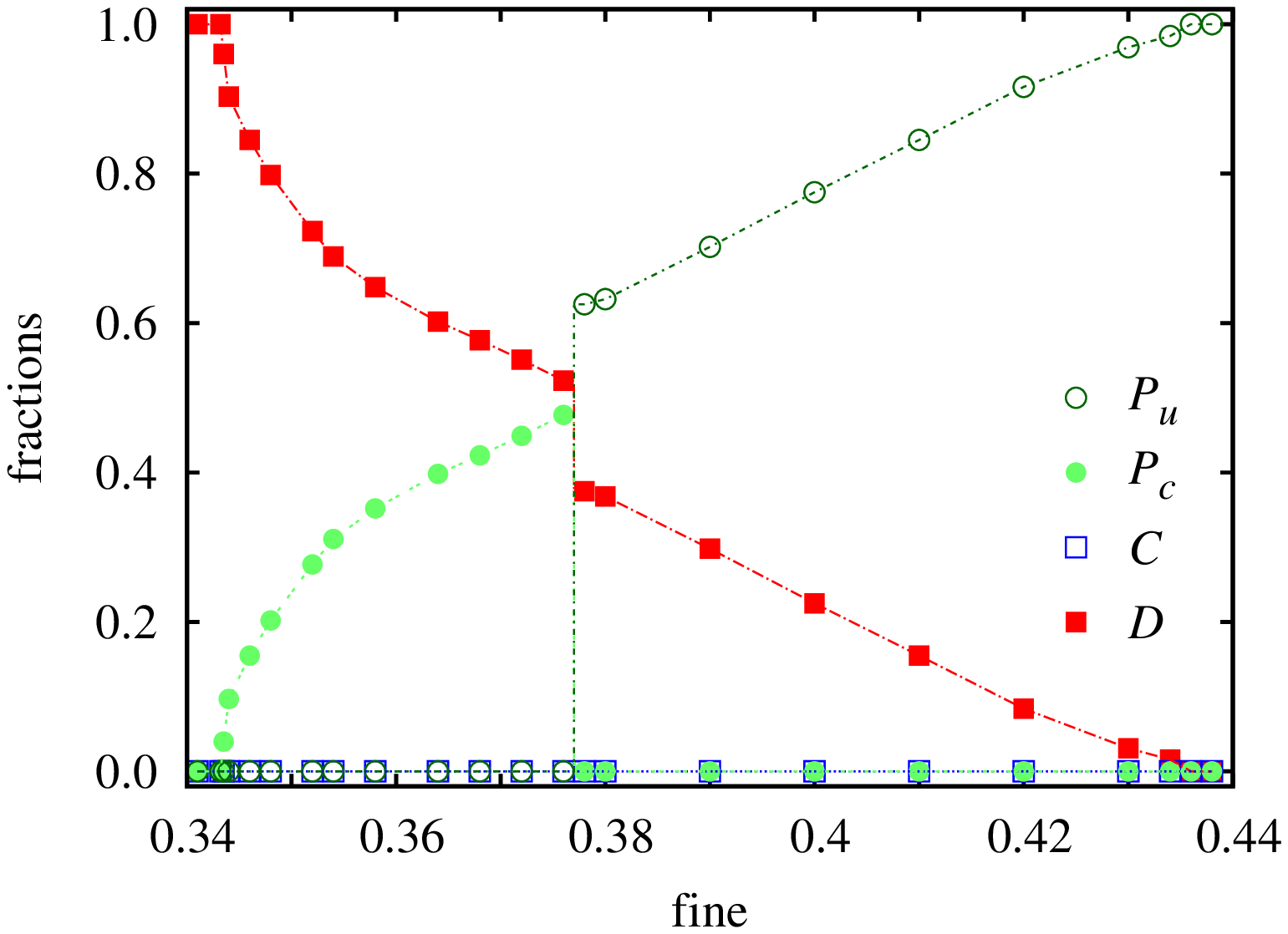,width=5.8cm}\epsfig{file=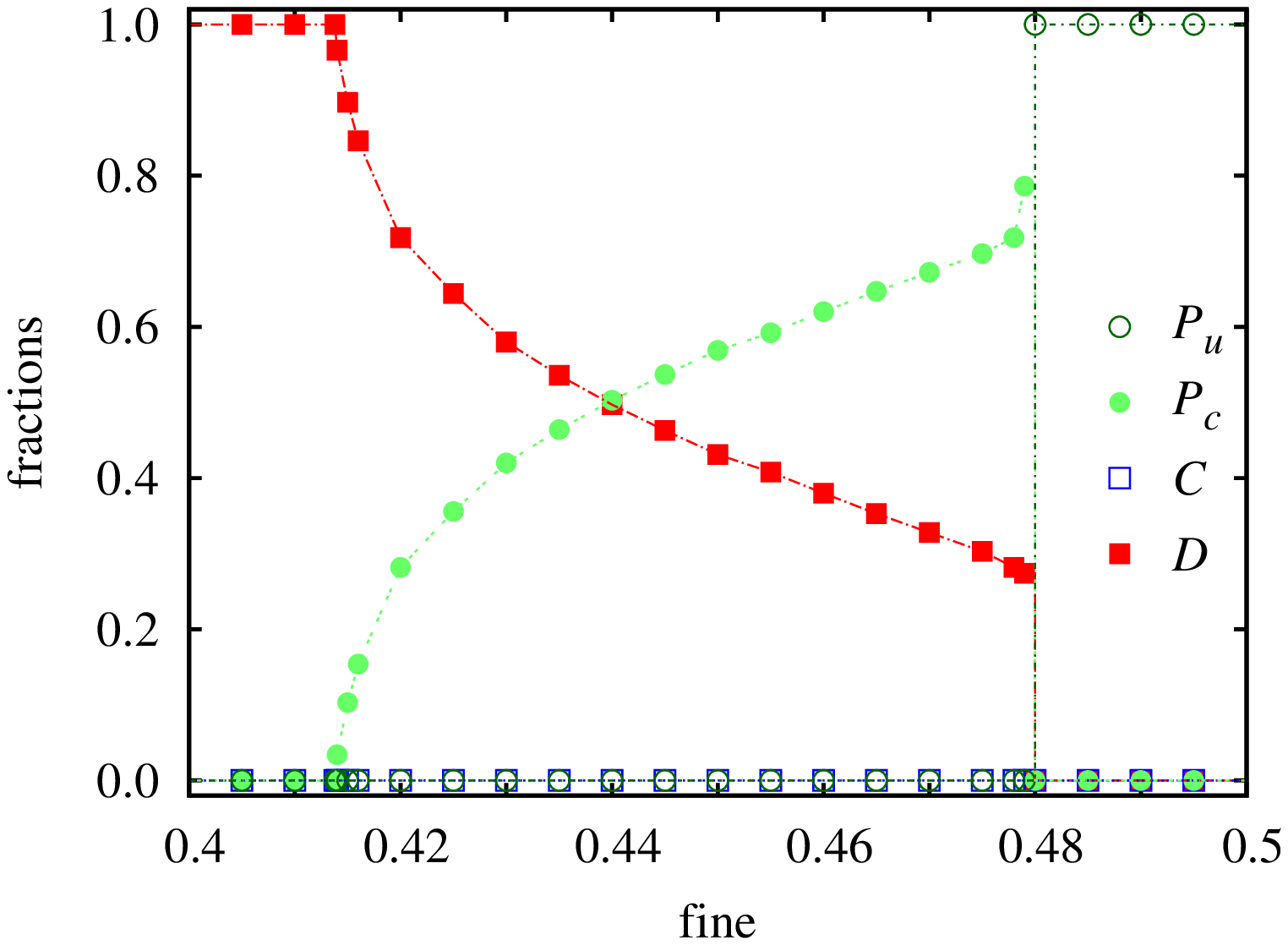,width=5.8cm}\epsfig{file=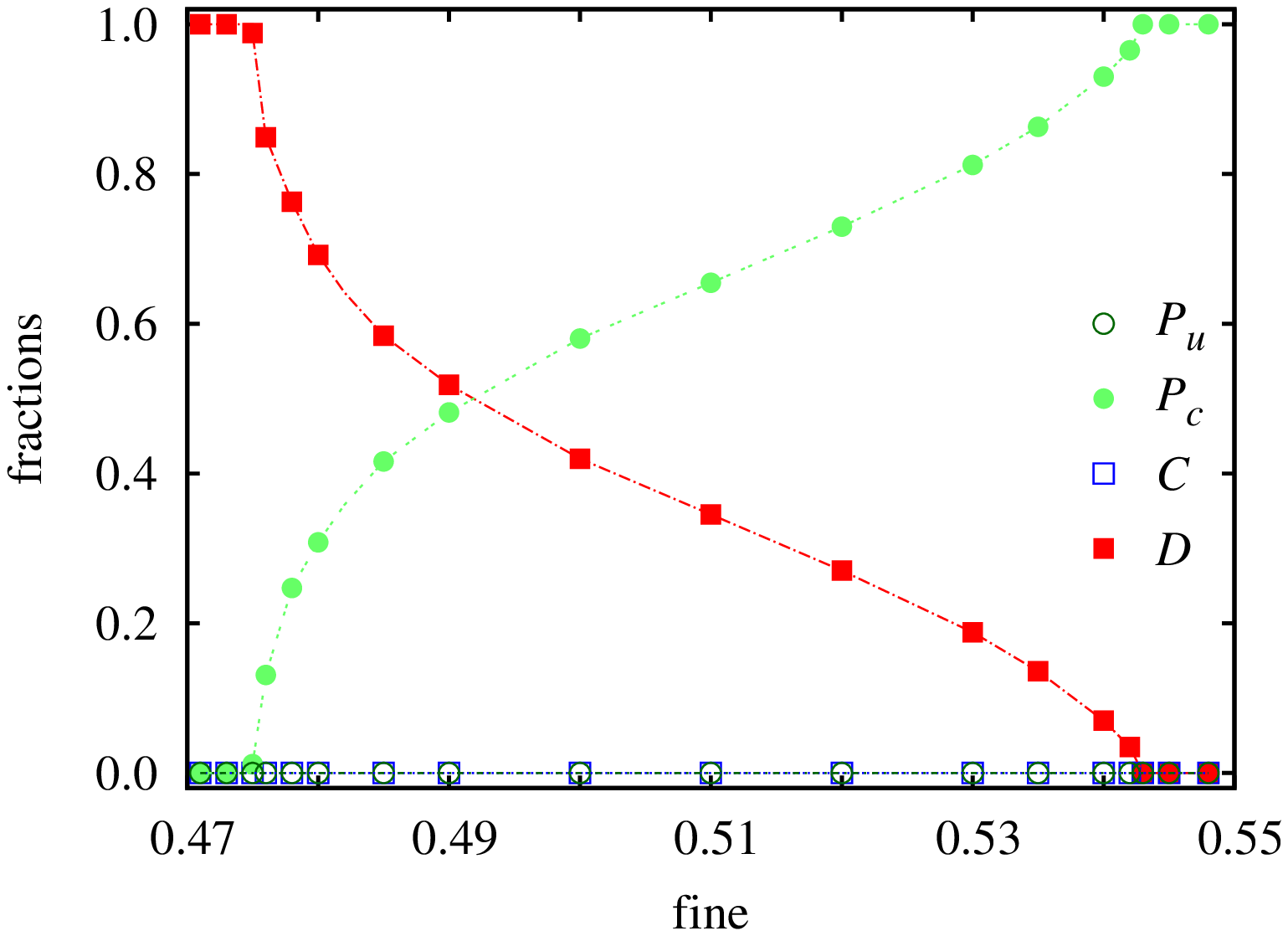,width=5.8cm}}
\caption{\label{cross35} Representative cross-sections of the phase diagram presented in Fig.~\ref{phase35}. Depicted are the stationary densities of the four competing strategies (see legend) in dependence on the punishment fine $\beta$, as obtained for three different values of cost: $\gamma=0.4$ (left), $\gamma=0.56$ (middle), and $\gamma=0.7$ (right).}
\end{figure*}

A more precise quantitative view of the evolutionary dynamics and the corresponding phase transitions can be obtained by means of representative cross-sections of the phase diagram, as presented in Fig.~\ref{cross38}. Left panel features the cross-section at $\gamma=0.3$, where as the punishment fine $\beta$ increases the discontinuous transition from the $D+C$ to the $D+P_c$ phase occurs first, followed by another discontinuous transition from the $D+P_c$ to the $D+P_u$ phase, which for even higher fines becomes the pure $P_u$ phase via a second-order continuous phase transition. The succession of phase transition at $\gamma=0.48$, still below the tricritical point, is slightly different in that the two-strategy $D+P_c$ phase transform directly into the absorbing $P_u$ phase, without the intermediate $D+P_u$ phase. The phase transition is discontinuous, and especially near such critical points a sufficiently large system size is of paramount importance. Here $P_u$ can easily become subject of accidental extinction if the system size is not large enough, and the seemingly stable solution in that case would appear to be the $D+P_c$ phase, which however would be a wrong result. In addition, the invasion of $P_u$ is extremely slow, frequently requiring more than $10^6$ full $MCS$ at $L=1600$ system size. If the punishment cost exceeds the tricritical point the succession of the phase transitions as $\beta$ increases changes yet again, as can be inferred from the right panel of Fig.~\ref{cross38}. In that case $P_u$ are unable to invade even at large values of $\beta$, and accordingly the mixed $D+P_c$ phase becomes the pure $P_c$ phase by means of a continuous phase transition.

If the multiplication factor $r$ is smaller than the threshold enabling the coexistence of cooperators and defectors, however, the phase diagram is topologically similar yet qualitatively different from the one presented in Fig.~\ref{phase38}. As can be observed in Fig.~\ref{phase35}, the mixed $D+C$ phase that characterized the low fine region at $r=3.8$ is missing. Instead, at $r=3.5$ we have a pure $D$ phase. Importantly, since cooperators die out, there is no indirect territorial competition between them and the punishers, which changes the nature of the phase transition line that marks the end of the pure $D$ phase. At sufficiently large fines and moderate punishment costs the pure $D$ phase becomes the mixed $D+P_u$ phase via a second-order continuous phase transition. Here unconditional punishers are able to take full advantage of the higher punishment fine and therefore outperform conditional punishers. As the punishment becomes more costly, however, the more economically acting conditional punishers become more efficient. The victor between $P_c$ and $P_u$ is again determined by means of an indirect territorial battle that is mediated by defectors. In particular, the punishing strategy that is more effective in resisting the invading defectors will ultimately share the space on the square lattice with them. Because of the indirect nature of the evolutionary competition, the phase transitions between the mixed $D+P_u$ and $D+P_c$ phases are discontinuous. The tricritical point above which unconditional punishers cannot survive, and where the discontinuous phase transition line merges with the continuous phase transition line, is for this value of the multiplication factor located at $\beta \simeq 0.53$ and $\gamma \simeq 0.63$. For high values of the punishment fine the evolutionary dynamics is the same as reported for $r=3.8$, in that the two punishing strategies become neutral as soon as defectors and cooperators die out, and the victor is thus determined by logarithmically slow coarsening during which the more widespread strategy is likelier to emerge as the dominant one.

\begin{figure*}[ht]
\centerline{\epsfig{file=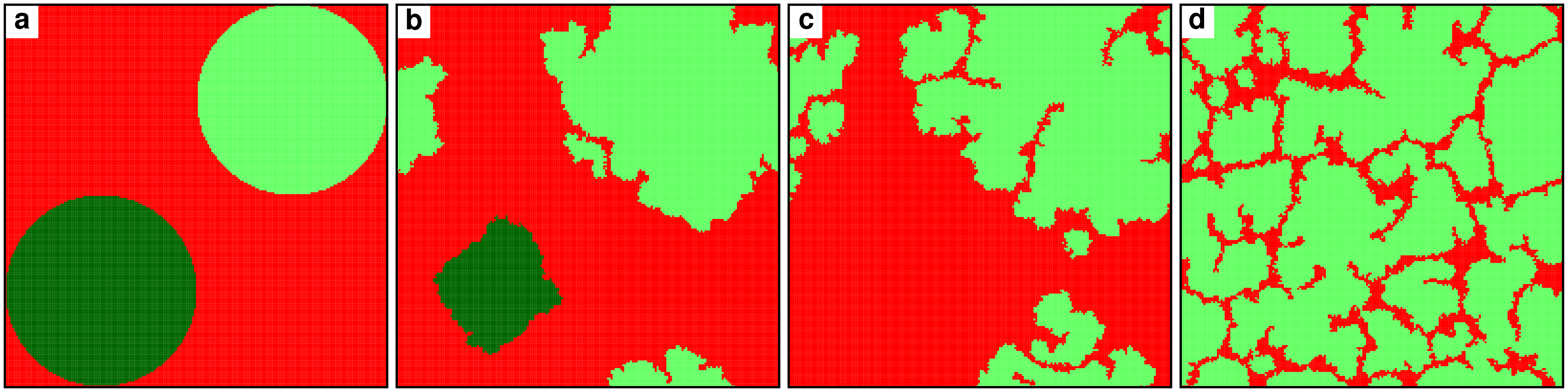,width=16cm}}
\centerline{\epsfig{file=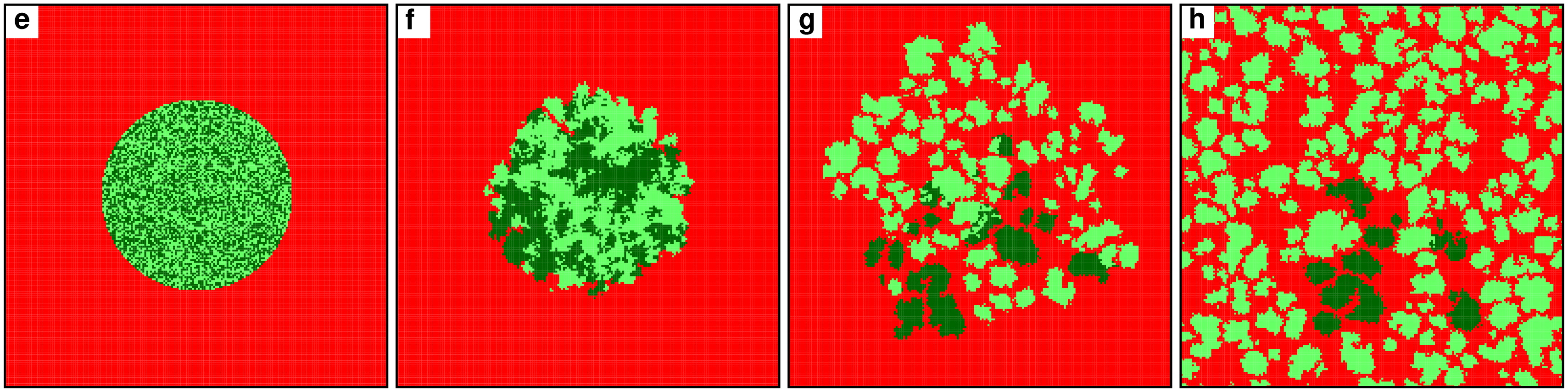,width=16cm}}
\caption{\label{snaps} Evolution of competing strategies from prepared initial states. Defectors are depicted red, while conditional and unconditional punishers are depicted light and dark green, respectively. Top row features the evolution of two initially isolated domains of conditional [upper right corner of panel (a)] and unconditional punishers [bottom left corner of panel (a)]  that are placed in the sea of defectors. Although unconditional punishers succeed in keeping a fully homogenous domain, the later shirks in size continuously [panel (b)], until it eventually vanishes completely [panel (c)].  Conditional punishers, on the other hand, allow ``cracks'' of defectors to emerge within their domain [panels (b) and (c)], yet still succeed in spreading and eventually forming a stable coexistence with the defectors [panel (d)]. Parameter values are $r=3.5$, $\beta=0.58$ and $\gamma=0.9$, while the snapshots were taken at $0$ (a), $1000$ (b), $2000$ (c) and $6000$ (d) full MCS. Bottom row features the evolution of a mixed domain consisting of conditional and unconditional punishers that is placed in the sea of defectors. Spontaneous coarsening of the two punishing strategies starts immediately [panel (b)], and soon they both form isolated domains that are surrounded by defectors [panel (c)]. From there on the evolutionary competition is determined by a relatively slow indirect territorial battle that is mediated by defectors [panel (d). The punishing strategy that is more successful against the defectors will ultimately prevail and form a stable coexistence with them. The less successful strategy, which in this particular case are the unconditional punishers, will die out (not shown). Parameter values are $r=3.5$, $\beta=0.37$ and $\gamma=0.4$, while the snapshots were taken at $0$ (a), $100$ (b), $1000$ (c) and $10000$ (d) full MCS. For clarity  the system size in all panels is $L=200$.}
\end{figure*}

Representative cross-sections of the phase diagram depicted in Fig.~\ref{phase35} are presented in Fig.~\ref{cross35}. In the left panel, obtained at $\gamma=0.4$, the continuous phase transition from the pure $D$ to the two-strategy $D+P_c$ phase occurs first as the fine $\beta$ increases. The $D+P_c$ phase then gives way to the $D+P_u$ via a first-order discontinuous phase transition, which is a consequence of the indirect territorial battle between $P_c$ and $P_u$ against defectors. For still higher values of $\beta$ the $D+P_u$ phase becomes the pure $P_u$ phase by means of a second-order continuous phase transition. For $\gamma=0.56$, depicted in the middle panel, the mixed $D+P_c$ transform directly into the pure $P_u$ phase, and here the same cautionary notes concerning the required system size and relaxation times are in order as issued above for $r=3.8$. Above the tricritical point, at $\gamma=0.7$ depicted in the right panel, unconditional punishers can no longer survive, and the evolutionary dynamics proceeds from the pure $D$ phase over the mixed $D+P_c$ to the pure $P_c$ phase by means of second-order continuous phase transitions only. Due to the absence of both cooperators and unconditional punishers this is indeed expected, as the indirect territorial competition is no longer possible.

To visualize and understand the leading mechanisms that are responsible for the reported evolutionary outcomes, it is instructive to study the evolution of spatial patterns from prepared initial states, as depicted in Figs.~\ref{snaps}(a) and (e). For clarity, we focus on $r=3.5$ and omit the initial presence of cooperators as they are indecisive for the composition of the final state. The described evolution of patterns is in fact generally valid and independent of $r$ as long as $\beta$ and $\gamma$ are adjusted to ensure the same stationary state.

We first focus on the parameter region where unconditional punishers are unable to survive. The evolution from a prepared initial state is depicted in the top row of Fig.~\ref{snaps}. It can be observed that the domain of unconditional punishers (dark green) remains completely homogeneous, yet is also shrinks in size continuously. Ultimately it vanishes, leading to the remainder of defectors (red) and conditional punishers (light green) as the only two competing strategies. Conditional punishers, on the other hand, proceed rather differently in the fight against defectors. Their less aggressive style of punishment allows small and narrow ``cracks'' of defectors to appear within the light green domain. Initially this may seem like a weakness, yet it turns out to be the winning recipe. Conceptually similar as reported recently for the public goods game on diluted lattices \citep{wang_z_pre12b} as well as for risk-driven migration \citep{chen_xj_pre12b}, such a configuration results in a sudden drop of public goods whenever defectors try to spread further. This in turn makes the defector strategy unlikely to be imitated further, and in fact the invasion is stopped. Although conditional punishers will in this way never be able to dominate the population completely, they do succeed in surviving alongside defectors at significantly lower fines than unconditional punishers, especially if the cost of sanctioning is comparable to the imposed fines, i.e., if the punishment is costly.

It is also instructive to examine the evolution from a differently prepared initial state, in the parameter region where both types of punishing strategies can in principle survive. The bottom row of Fig.~\ref{snaps} depicts the evolution of a mixed $P_c+P_u$ domain in the sea of defectors. Practically instantly, after only 100 MCS, the two punishing strategies start coarsening, eventually forming compact isolated domains that are surrounded by defectors. This evolution demonstrates nicely that the second-order exploitation that was raised in several well-mixed solutions is not necessarily viable. Realistically, the interactions we have with others are always restricted. Everybody is not connected to everybody else, not even on average and neither in the long run. Given the restricted neighborhoods, some sort of coarsening will always happen due to imitation, even if the strategies are neutral. Consequently, smaller communities may become homogeneous, and they may proceed with their competition against a certain strategy, yet independently of other strategies that may also be present in the population at the time. As the panel (g) of Fig.~\ref{snaps} shows, locally this process will not necessarily result in the victory of the more efficient strategy. Besides light green islands denoting conditional punishers, dark green islands denoting unconditional punishers are formed too. More to the point, the $D+P_u$ phase would actually be stable, were it not for the presence of $P_c$, who will eventually crowd out $P_u$ by means of the indirect territorial battle with the defectors. At this point we again emphasize the apparent general applicability of indirect territorial competition as a mechanism by means of which seemingly subordinate strategies may turn out to be evolutionary stable and prevail over the superior ones. It is exactly this mechanism that allows punishers, despite their obvious disadvantage over second-order free-riders, to nevertheless prevail in a structured population without any additional incentives or strategic complexity \citep{helbing_ploscb10}, and it is the same mechanism that allows conditional punishers to prevail over unconditional punishers despite their inherently less aggressive style of administering the fines to defectors.

\begin{figure}[h!]
\centerline{\epsfig{file=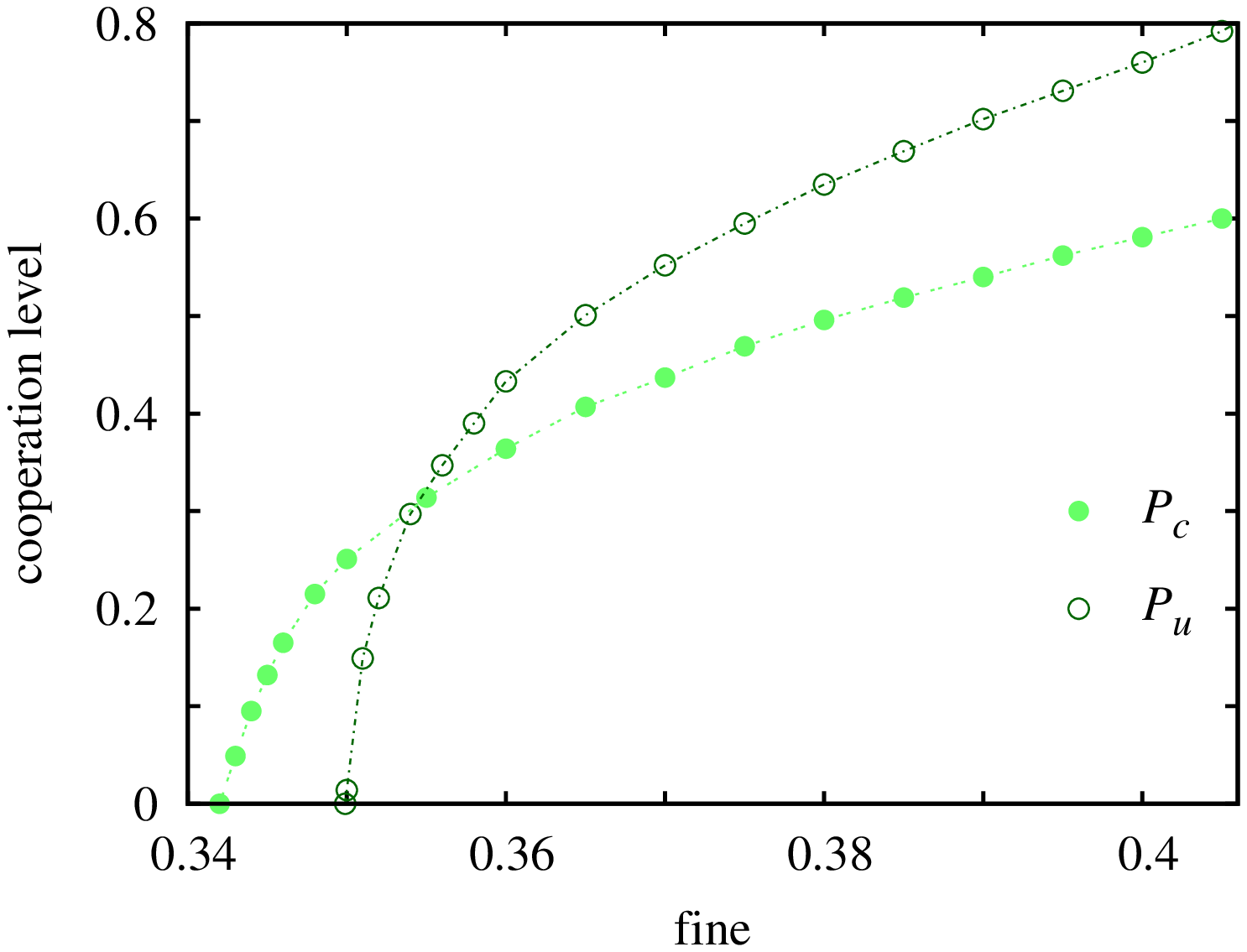,width=8cm}}
\centerline{\epsfig{file=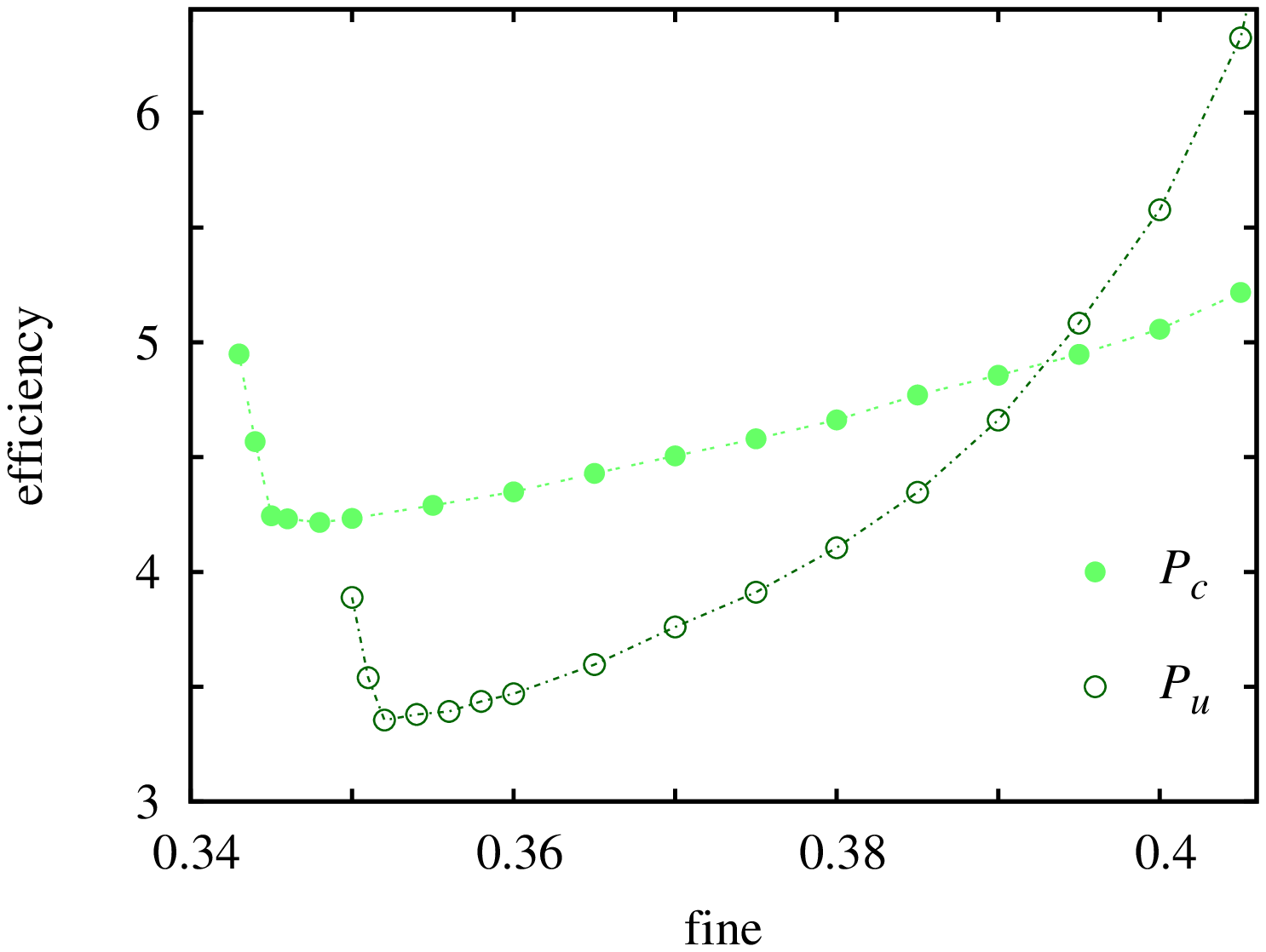,width=8cm}}
\caption{\label{compare} Comparison of efficiency of conditional and unconditional punishment in two three-strategy public goods games, entailing besides the traditional defectors and cooperators also conditional (filled light green circles) and unconditional punishers (open dark green circles), respectively. Top panel shows the cooperation level in dependence on fine at $\gamma=0.4$ and $r=3.5$. Bottom panel depicts the corresponding efficiency of punishment in the two games, defined as the ratio between the cooperation level and the average cost that is necessary to maintain it. Conditional punishment is more effective, if only the punishment is not excessively cheap.}
\end{figure}

Since the question of effectiveness of conditional versus unconditional punishment is far from trivial given that the ratio between cost and fine is always the same, it is lastly informative to compare their relations directly in a quantitative manner. To do so properly, we compare the efficiency of two three-strategy games, namely the public goods game entailing just unconditional punishers, as studied previously in \citep{helbing_njp10}, and the public goods game entailing just conditional punishers. By focusing on the relevant parameter region where the two punishing strategies can independently coexist with defectors, we plot in the top panel of Fig.~\ref{compare} the cooperation level, i.e., the fraction of $P_c$ and $P_u$ (note that $C$ die out due to small $r$), as the function of fine. It can be observed that conditional punishment lowers the threshold value of fine $\beta$ at which punishers can grab a hold in the population. On the other hand, for larger values of $\beta$ the fraction of $P_u$ increases fast and quite quickly exceeds that of $P_c$. This invites the conclusion that indeed the unconditional punishment might be more effective, at least indirectly. Yet this is in general not true. If the punishment is costly the efficiency of conditional punishment is larger, as can be demonstrated clearly if we normalize the cooperation level by the average cost that is necessary to maintain it. The bottom panel of Fig.~\ref{compare} features the result, which evidences that the efficiency is almost always higher for conditional punishment, except when the punishment becomes really cheap. This difference also explains why $D+P_c$ can prevail over $D+P_u$ for lower fines.

\section{Discussion}
We have studied the effectiveness of conditional punishment in promoting public cooperation, in particular comparing it to the effectiveness of the more commonly considered unconditional punishment. We have shown that in the four-strategy public goods game entailing cooperators and defectors as well as conditional and unconditional punishers, the later two strategies cannot coexist. Spontaneous coarsening leads to their segregation on the spatial grid, upon which they compete against each other indirectly through their rivalry with defectors. If punishment is cheap, i.e., if either the cost of punishing is low or the fine is comparatively large, unconditional punishers are more effective in invading defectors, which in turn crowds out conditional punishers. Conversely, in the more realistic case when the punishment is costly, conditional punishers are more successful in deterring defectors, which leads to the extinction of unconditional punishers. For sufficiently large fines, however, defectors die out completely, which makes the two punishing strategies equivalent, and the victor between them is determined by means of logarithmically slow coarsening, as is known from the voter model \citep{dornic_prl01}. Details of these evolutionary relations, however, depend somewhat also on the multiplication factor $r$. If the later is sufficiently large so that cooperators can survive alongside defectors even in the absence of punishment, then the mixed $D+C$ phase first gives way to the mixed $D+P_c$ phase via a first-order discontinuous phase transition. In this case cooperators and conditional punishers compete against each other indirectly through defectors. If the multiplication factor is lower, on the other hand, the pure $D$ phase becomes either the $D+P_u$ or the $D+P_c$ phase through a second-order continuous phase transition, depending on the punishment cost. The ubiquity of indirect territorial competition in the public goods game with conditional punishment generalizes the observations of our previous work \citep{helbing_ploscb10}, where such evolutionary dynamics was reported first between $D+C$ and $D+P$, where $P$ were considered to be unconditional punishers. Here we show that it may emerge also between $D+C$ and $D+P_c$ as well as between $D+P_u$ and $D+P_c$, and in all cases it leads to discontinuous phase transitions, which under special conditions may transform into continuous phase transitions via a tricritical point in the corresponding phase diagram. We argue that indirect territorial competition constitutes a general mechanism that is driven by pattern formation, by means of which spatial structure can be exploited to create evolutionary advantages for strategies that are obviously inferior in well-mixed populations. Notably, the absence of such complex evolutionary scenarios in traditional physics systems is due to the multi-point interactions that emerge because of conditional punishment, which in turn enriches not only our understanding of the evolution of public cooperation, but also reveals new ways by means of which pattern formation can manifest itself in interacting particle systems \citep{liggett_85}.

In general, the larger efficiency of conditional punishment to sustain cooperation in the face of defection lies, quite paradoxically, in the lesser efficiency of conditional punishers to grow and maintain completely compact homogeneous clusters. Although this prohibits the total extinction of defectors, it also enables the spreading of conditional punishers. The effect is conceptually similar as reported recently for diluted lattices \citep{wang_z_pre12b} and risk-driven migration \citep{chen_xj_pre12b}, where it was shown that ``cracks'' in the otherwise compact cooperative domains lead to a sudden drop of public goods whenever defectors try to spread further. This in turn makes the defector strategy less attractive for the neighbors, and indeed the invasion via imitation is thereby stopped. It is worth noting on this occasion that the evolutionary advantages of imitation, even at weak selection pressure, are hardly disputable \citep{masuda_srep12, mobilia_pre12, szolnoki_srep12}. In our particular case the relatively mild application of punishment as administrated by conditional punishers, along with the relatively lower cost, turns out to be the more effective cure against the invading defectors than hard unconditional punishment. Conditional punishers do allow a relatively small fraction of defectors to survive inside cobweb-like cracks that are spread across the spatial grid, yet this seeming weakness in fact forms the backbone of their deceptively simple yet very effective protection against further invasions. This is also why conditional punishers are able to maintain cooperation at lower fines than unconditional punishers, and why the efficiency of the former is in general higher. Exceptions are parameter regions where punishment is really cheap, which are evolutionary less interesting and in fact trivial due to a fully predictable final outcome. Nevertheless, the message is if the execution of the penalty is cheap, it may as well be a strong one. Under more realistic circumstances, where the expenses of punishment need to be taken into account, however, it always makes more sense to punish conditionally.

\section*{Acknowledgments}
This research was supported by the Hungarian National Research Fund (Grant K-101490) and the Slovenian Research Agency (Grant J1-4055).


\end{document}